\documentclass[twocolumn,floats,aps,superscriptaddress]{revtex4}
\usepackage{times} 
\usepackage{amssymb,amsmath}
\ifx\pdfoutput\undefined
\usepackage[usenames,dvips]{color}
\usepackage{graphicx}     
\else
\usepackage[usenames,dvipsnames]{color}
\usepackage{epstopdf}
\usepackage[pdftex]{graphicx}
\usepackage[pdftex]{hyperref}
\hypersetup{a4paper,
    pdftitle={Manuscript on XXX} 
    pdfauthor={Uwe Thiele},  
pdfproducer={lateX},
pdfview=FitV,       
pdfstartview=FitB,
linkcolor=blue,     
citecolor=blue,     
pagecolor=blue,     
urlcolor=red,      
breaklinks=true,    
colorlinks=true,
citebordercolor=0 0 0,  
filebordercolor=0 0 0,
linkbordercolor=0 0 0,
menubordercolor=0 0 0,
pagebordercolor=0 0 0,
urlbordercolor=0 0 0,
pdfhighlight=/I,
pdfborder=0 0 0,   
backref=false,
pagebackref=false,
bookmarks=true,
bookmarksopen=true,
bookmarksnumbered=true
}
\fi
\ifx\pdfoutput\undefined
\DeclareGraphicsExtensions{.eps}
\else
\DeclareGraphicsExtensions{.jpg, .pdf, .tif, .png}
\fi
\graphicspath{{.},{./Figures/}} 
\usepackage[usenames,dvipsnames]{color}
\begin{document}
\title{Continuous and discontinuous dynamic unbinding transitions in drawn film flow}
\author{M. Galvagno}
\author{D. Tseluiko}
\affiliation{Department of Mathematical Sciences, Loughborough University,
Loughborough, Leicestershire, LE11 3TU, UK}
\author{H. Lopez}
\affiliation{School of Physics and Complex and Adaptive Systems Laboratory,
University College Dublin, Belfield, Ireland}
\author{U. Thiele}
\affiliation{Department of Mathematical Sciences, Loughborough University,
Loughborough, Leicestershire, LE11 3TU, UK}
\affiliation{Institut f\"ur Theoretische Physik, Westf\"alische
 Wilhelms-Universit\"at M\"unster, Wilhelm Klemm Str.\ 9, D-48149 M\"unster, Germany}
%
\begin{abstract}
  When a plate is withdrawn from a liquid bath a coating layer is
  deposited whose thickness and homogeneity depend on the velocity and
  the wetting properties of the plate. Using a long-wave mesoscopic
  hydrodynamic description that incorporates wettability via a
  Derjaguin (disjoining) pressure we identify four qualitatively
  different dynamic transitions between microscopic and macroscopic
  coatings that are out-of-equilibrium equivalents of well known
  equilibrium unbinding transitions. Namely, these are continuous and
  discontinuous dynamic emptying transitions and discontinuous and
  continuous dynamic wetting transitions. We uncover several features
  that have no equivalent at equilibrium.
\end{abstract}
\maketitle

The equilibrium and non-equilibrium behaviour of mesoscopic and
macroscopic drops, meniscii and films of liquid in contact with static
or moving solid substrates is not only of fundamental interest but
also crucial for a large number of modern technologies.  On the one
hand, the equilibrium behaviour of films, drops and meniscii is
studied by means of statistical physics. A rich substrate-induced
phase transition behaviour is described even for simple liquids, e.g.,
related to wetting and emptying transitions that both represent
unbinding transitions. In the former case the thickness of an
adsorption layer of liquid diverges continuously or discontinuously at
a critical temperature or strength of substrate-liquid interaction,
i.e., the liquid-gas interface of the film unbinds from the
liquid-solid interface \cite{BEIM2009rmp}. In the case of the emptying
transition a macroscopic meniscus in a tilted slit capillary develops
a tongue (or foot) along the lower wall of a length that diverges
logarithmically at a critical slit width, i.e., the tip of the foot
unbinds from the meniscus \cite{PRJA2012prl}.

\begin{figure}[htbp!]
\includegraphics[width=1\hsize]{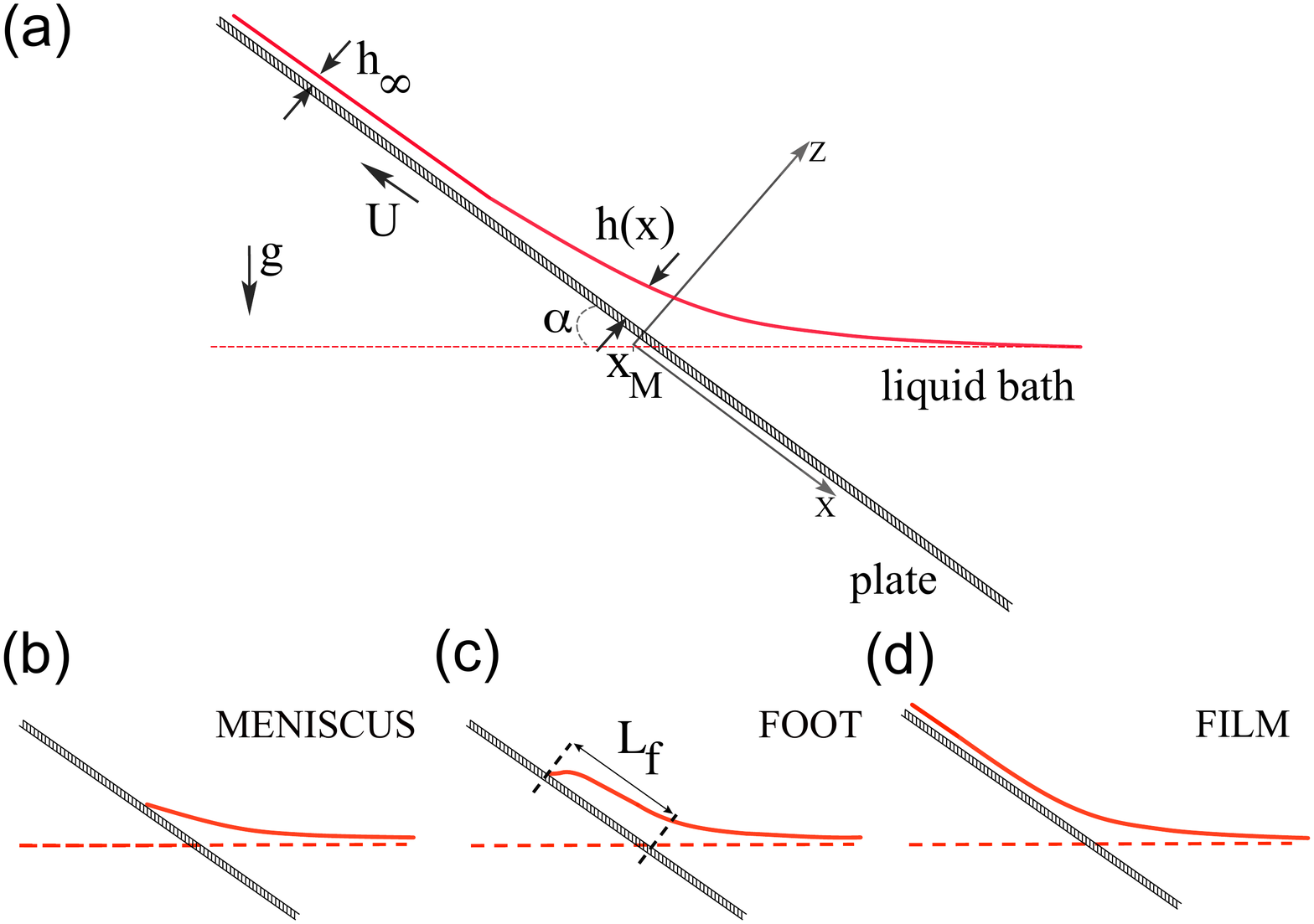}
\caption[Solution types]{Sketches of (a) the considered
  two-dimensional geometry and (b-d) of the qualitatively different steady
  shapes $h(x)$ of the free liquid surface as found in experiments:
  (b) {\it simple meniscus}, (c) {\it foot} or {\it extended meniscus}, and
  (d) {\it Landau-Levich film}. In (a) a flat plate inclined at an
  angle $\alpha$ to the horizontal is drawn out of a bath of a
  non-volatile partially wetting liquid at a speed $U$. The introduced
  Cartesian coordinate system has its $x$-axis parallel to the plate
  and its origin fixed to the point where an ideally horizontal bath
  surface would meet the plate.}
\label{fig:dragged_fig}
\end{figure}%

On the other hand, it is a classical hydrodynamic problem to study how
droplets slide down an incline
\cite{PFL2001prl,TNBP2002csapea,BCP2003pf,SnAn2013arfm}, how moving
contact lines (where solid, gas and liquid meet) develop sawtooth
shapes at high speeds \cite{BlRu1979n,DFSA2008jfm,SnAn2013arfm}, or
how the free surface of a bath is deformed when a plate is drawn out,
as sketched in Fig.~\ref{fig:dragged_fig}(a). Early on it was reported
that for sufficiently large plate velocities $U$ a homogeneous
macroscopic liquid layer is deposited on the drawn plate
[Fig.~\ref{fig:dragged_fig}(d)].  The resulting coating layer is
called a Landau-Levich film. Far away from the bath it has a thickness
$h_\infty$ that depends on the capillary number $\mathrm{Ca}=\eta
U/\gamma$ through the power law $h_\infty\propto\mathrm{Ca}^{2/3}$
\cite{LaLe1942apu} where $\eta$ and $\gamma$ are the viscosity and
surface tension of the liquid, respectively. This coating technique is
widely used and became a paradigm for theoretical (e.g.,
\cite{derj1943apu,LaLe1942apu,Wils1982jem,MuEv2005pd,SnAn2013arfm})
and experimental (e.g.,
\cite{gowa1922pm,groe1970ces,SDFA06prl,SZAF2008prl,MRRQ2011jcis})
studies.

In contrast, at very low plate velocities $U$ no macroscopic film is drawn
out but a deformed steady meniscus coexists with the dry plate far
away from the bath
\cite{Egge2004prl,SADF2007jfm,DFSA2008jfm,SZAF2008prl,ChSE12pf}
[Fig.~\ref{fig:dragged_fig}(b)]. This meniscus only exists for
capillary numbers smaller than a critical one, i.e.,
$\mathrm{Ca}<\mathrm{Ca}_c$ \cite{Egge2004prl}.  Close to
$\mathrm{Ca}_c$ the meniscus develops a foot of a length $L_f$
[Fig.~\ref{fig:dragged_fig}(c)] that diverges at $\mathrm{Ca}_c$
either continously \cite{ZiSE2009epjt} or discontinuously
\cite{SADF2007jfm}.  As the steady free surface meniscus coexists with
the dry moving plate, there exists a receding three-phase contact line
whose best description is still debated (see, e.g.,
\cite{deGe1985rmp,BEIM2009rmp,Vela2011epjst}).

Previous works \cite{Egge2004prl,SADF2007jfm,ZiSE2009epjt} employ a
slip model that allows the film height to go to zero at the contact
line and avoids the contact line singularity through the slip
\cite{BEIM2009rmp}. Although, a slip model allows for a quantitative
study of meniscus solutions and Landau-Levich films, it is not able to
describe transitions between them as in a slip model they are
topologically different [cf.~Figs.~\ref{fig:dragged_fig}(b) and~(c)
vs.~Fig.~\ref{fig:dragged_fig}(d)]. Note that this concerns the actual
transition dynamics as well as the description of transitions in
dependence of control parameters such as the plate speed.

In contrast, here we employ a long-wave mesoscopic hydrodynamic model
that describes wettability via a Derjaguin (disjoining) pressure,
i.e., a precursor film model.  An investigation of the non-equilibrium
transitions between meniscus solutions and Landau-Levich films then
allows for an identification of four qualitatively different dynamic
unbinding transitions, namely, continuous and discontinuous dynamic
emptying transitions and discontinuous and continuous dynamic wetting
transitions. Note that far from the transition regions, the
predictions of precursor and slip models agree very well and can be
quantitatively mapped \cite{SaKa2011el}.

In particular, to describe the meniscus and the film dynamics we use
the following non-dimensionalised \cite{noteScaling} evolution
equation for the film thickness profile $h(x,t)$
~\cite{Thie2007,deGe1985rmp,Thie2010jpcm}:
\begin{eqnarray}
\partial_t h&=&-\partial_x\left\{ h^3 \partial_x [\partial_x^2h+\Pi(h)]\right.\nonumber\\
&&\left.\qquad\qquad-h^3 G(\partial_x h-\alpha)-U h\right\},
\label{eq:thifi_dim_non}
\end{eqnarray}
that may be derived as long--wave approximation of the Navier-Stokes
and continuity equations with no-slip boundary conditions at the
liquid-solid interface and kinematic and stress balance conditions at
the liquid-gas interface \cite{OrDB1997rmp}. Here $U$, $G$ and
$\alpha$ are the scaled plate velocity (Capillary number), gravity
(Bond number), and inclination angle of the plate, respectively
\cite{noteScaling}.  The wettability of the partially wetting liquid
is described via the Derjaguin (or disjoining) pressure \cite{StVe2009jpcm}
 \begin{equation}
\Pi=-\frac{\mathrm{1}}{h^3} \left(1 - \frac{1}{h^3}\right),
\label{eq:derja}
\end{equation} 
derived in Ref.~\cite{Pism2002csaea} from a modified Lennard-Jones
potential with hard-core repulsion, see \cite{noteScaling}.  The
disjoining pressure is related to a wetting or adhesion energy $f(h)$
via $\Pi=-df/dh$.

\begin{figure}[htbp!]
\includegraphics[width=0.49\hsize]{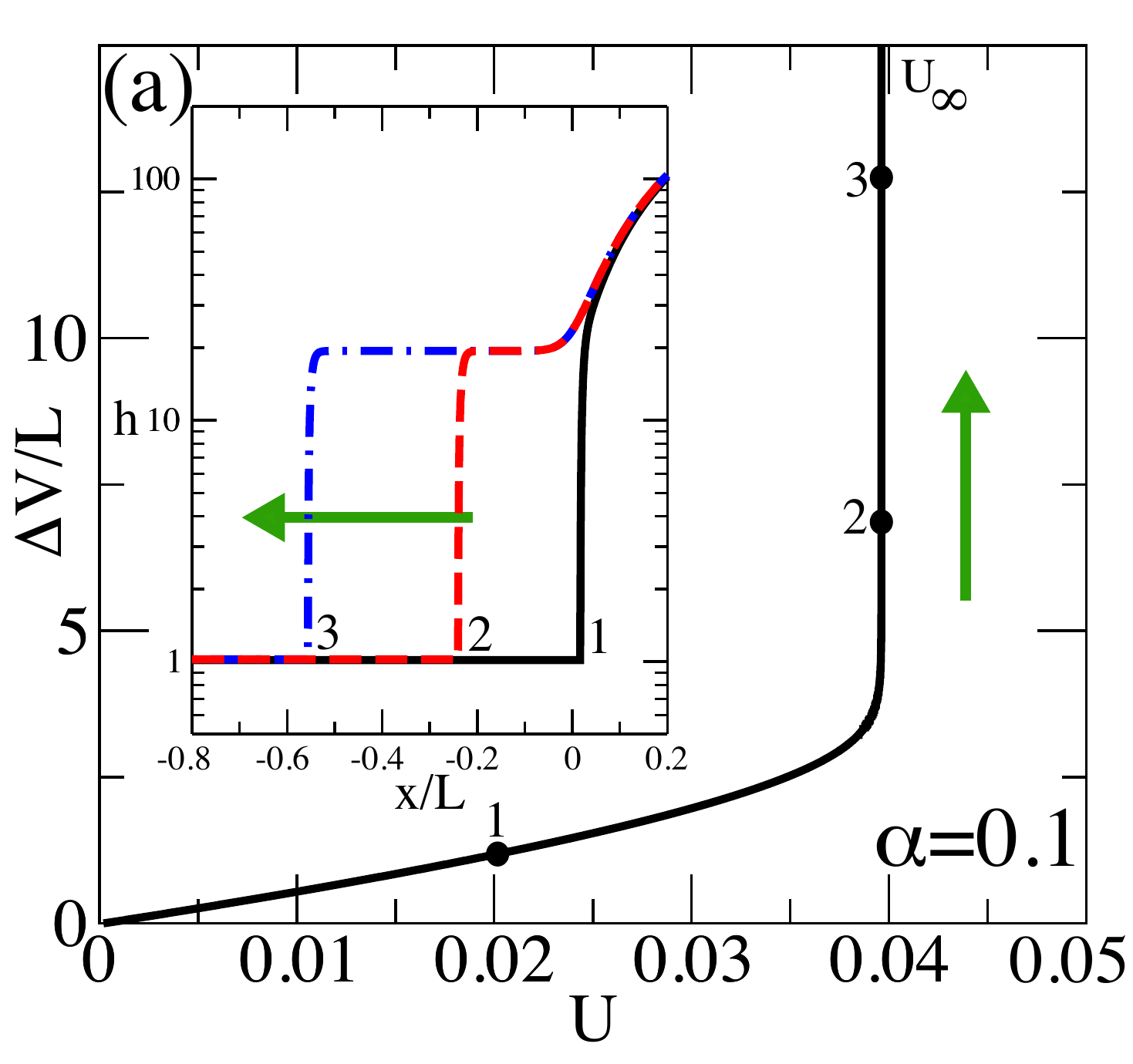}
\includegraphics[width=0.49\hsize]{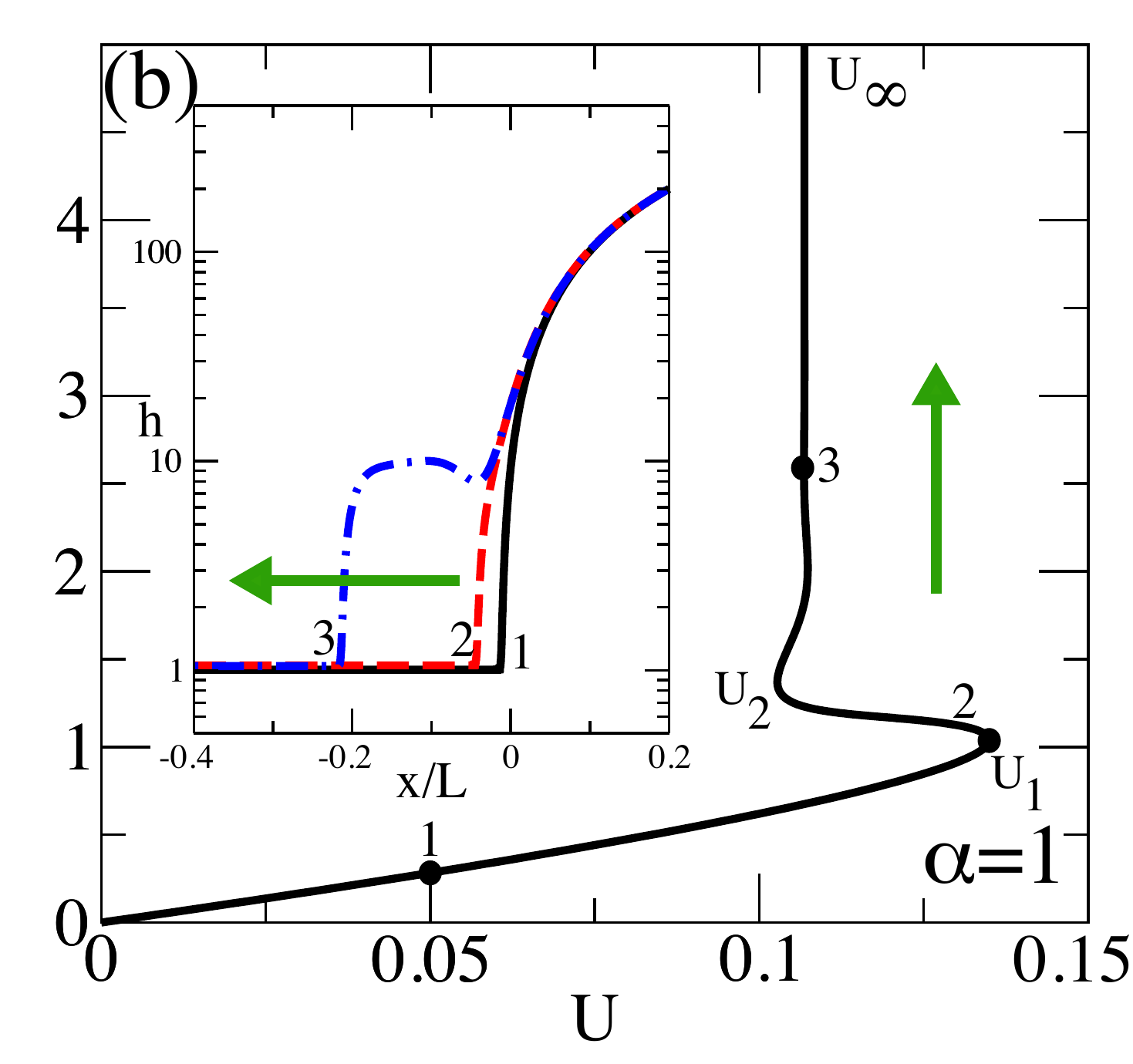}
\includegraphics[width=0.49\hsize]{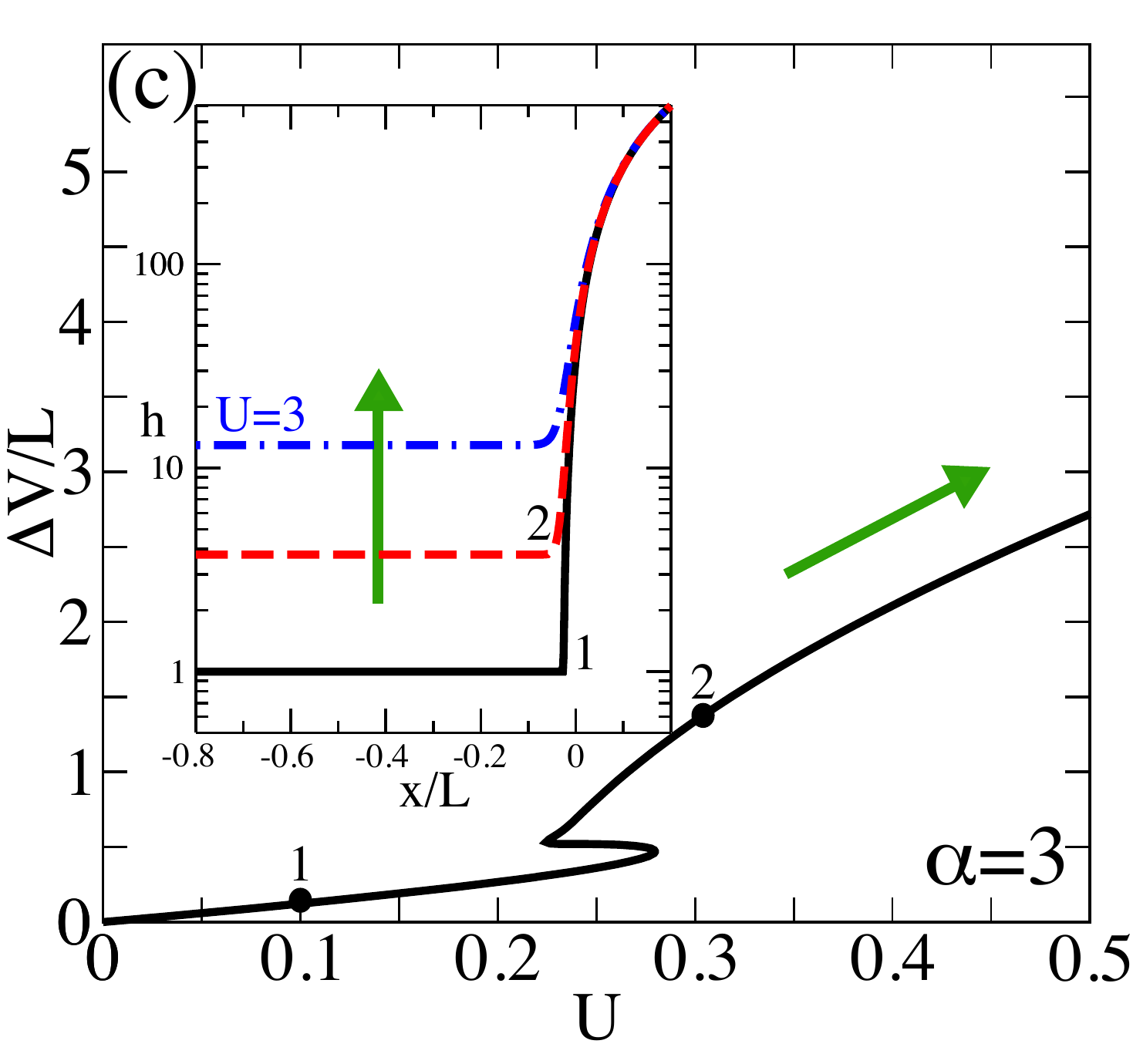}
\includegraphics[width=0.49\hsize]{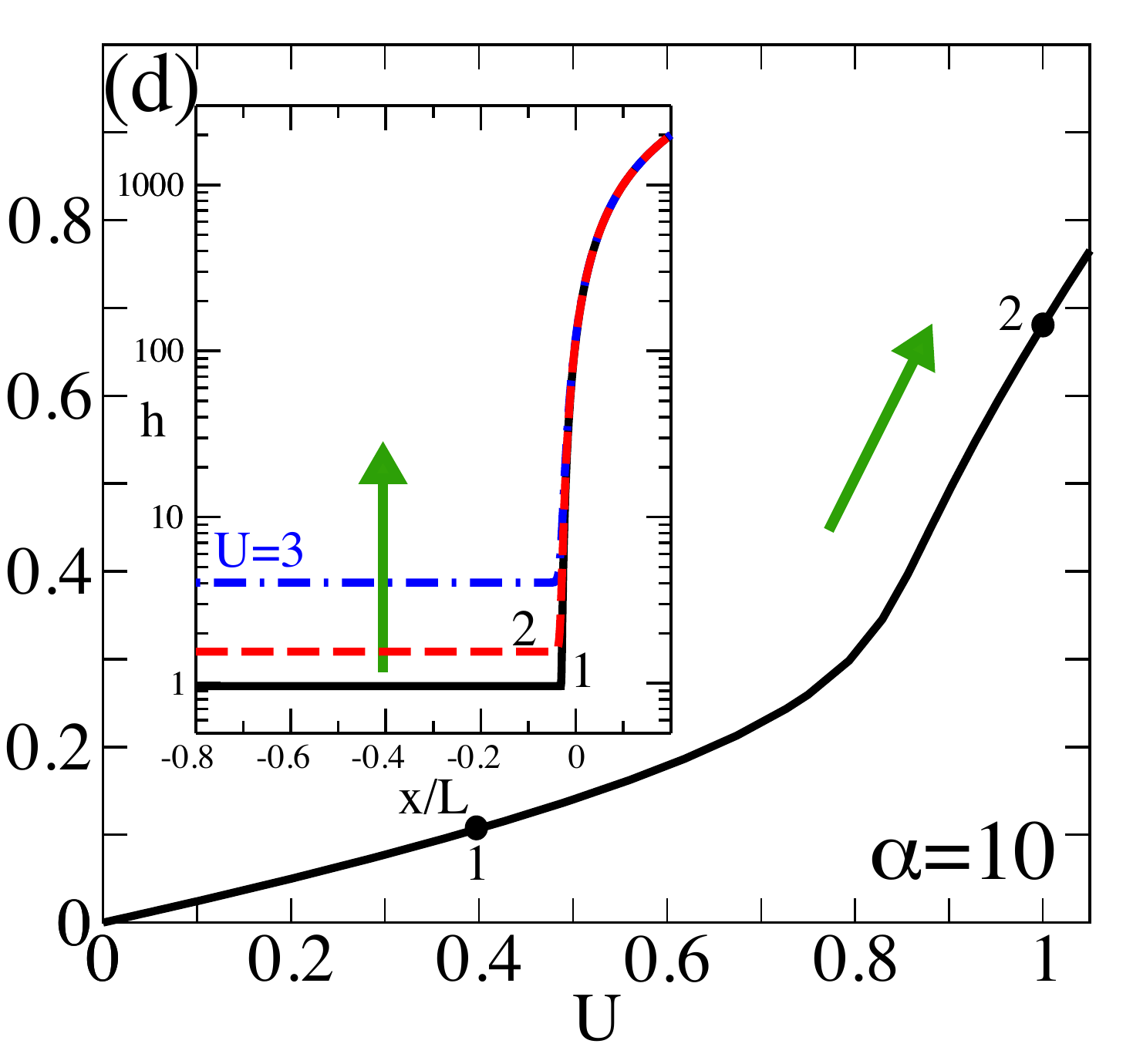}
\caption{Bifurcation curves indicating the occurence of qualitatively different
  behaviour with increasing plate inclination angles (a) $\alpha = 0.1$,  (b)
  $\alpha = 1$,  (c) $\alpha = 3$, and (d) $\alpha = 10$.
The main panels shown the excess
volume $\Delta V$ over domain size $L$ (see main text) in dependence of the plate
  velocity $U$, while the
  respective insets give Log-normal representations of steady film
  profiles at selected plate velocities as indicated by corresponding
  labels at the profiles and at the bifurcation curves. Additionally,
  panels (c) and (d) give a film profile at 
$U = 3$. The domain size is $L = 1000$. Arrows indicate how the profiles
  change as one moves along the bifurcation curves.}
\label{fig:profiles}
\end{figure}

It should be noted that the hydrodynamic long-wave model,
Eq.~(\ref{eq:thifi_dim_non}) with Eq.~(\ref{eq:derja}), directly
corresponds to a gradient dynamics of an underlying interface
Hamiltonian (or free energy) $F[h]=\int [\xi\gamma+f(h)]dx$ 
as often
used to study the above introduced equilibrium unbinding transitions
\cite{notehamiltonian}. This equivalence allows for a natural
understanding of the various occuring transitions as non-equilibrium
(or dynamic) unbinding transitions (see below).

\begin{figure}
\includegraphics[width=1.0\hsize]{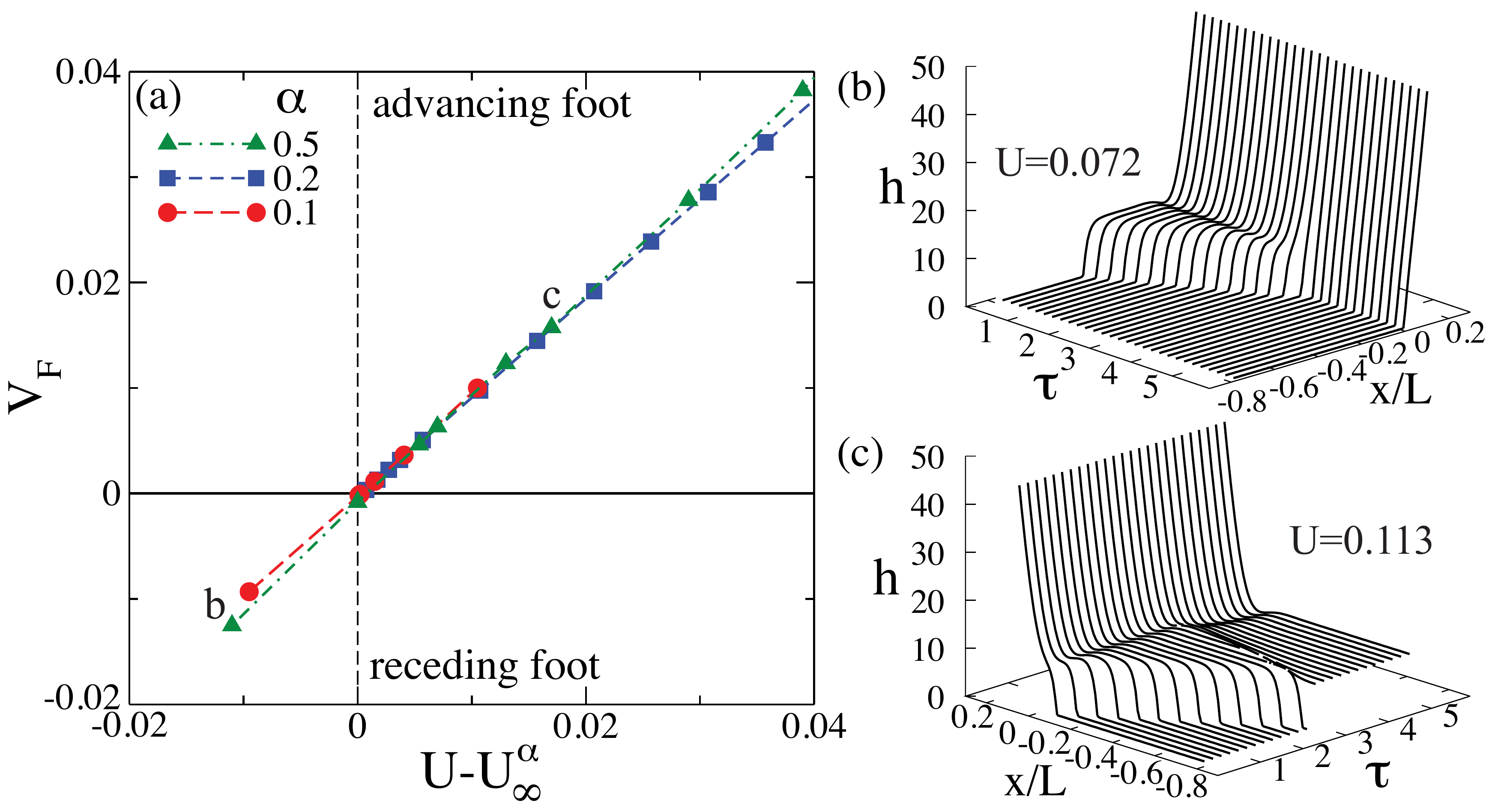}
\caption{(a) Advancing and receding foot-like structures are
  characterized by the dependence of the velocity $V_\mathrm{F}$ of
  the front that connects the ultrathin coating layer of thickness
  $h_\infty$ with the foot plateau of height $h_\mathrm{foot}$ on the
  velocity difference $U-U_\infty^{\alpha}$ where $U_\infty^{\alpha}$
  changes with the plate inclination $\alpha$. Note that the curves
  for various $\alpha$ as given in the legend collapse onto a master
  curve, indeed $V_\mathrm{F}\approx U-U_\infty^{\alpha}$.  Panels (b)
  and (c) give for $\alpha=0.5$ space-time plots representing the time
  evolution \cite{noteSim} of a receding and an advancing foot, respectively, at values
  of $U$ indicated by small letters in panel (a). The
  evolution in (b) converges to a steady simple meniscus, while in (c)
  the foot advances with constant speed until its tip reaches the
  domain boundary. Then at $\tau\approx4$ the foot
  transforms into a Landau-Levich film of a different thickness via a
  fast shallow backwards-moving front.}
\label{fig:frontvel}
\end{figure}

To calculate steady film and meniscus profiles one sets $\partial_t
h=0$, then integrates Eq.~(\ref{eq:thifi_dim_non}) once and solves the
resulting three-dimensional dynamical system in $(h,\partial_x
h,\partial_{xx} h)$ with appropriate boundary conditions: (i) far from
the bath for $x\rightarrow-\infty$ one imposes that the film profile
approaches a flat film of unknown height $h_\infty$ that is determined
as part of the solution while (ii) the approach towards the bath for
$x\rightarrow\infty$ is described by an asymptotic series that can be
rigorously derived via a center manifold reduction
\cite{TsGT2013arxiv}.  Steady profiles and bifurcation diagrams are
numerically obtained employing pseudo-arclength continuation
\cite{DWCD2014ccp}.  The employed main solution measure is the dynamic
excess volume $\Delta V= V -V_0$ with $V=\int (h(x)-h_\infty)dx$,
where $V_0$ is $V$ at $U=0$. Note that for solutions with a long
protruding foot-like structure $\Delta V$ is approximately
proportional to the length of the foot \cite{notefoot}.

An analysis of the changes that steady meniscii undergo with
increasing plate speed $U$ shows that four qualitatively different
cases exist depending on the plate
inclination angle $\alpha$.
Each case is related to a distinguished nonequilibrium unbinding
transition as illustrated in Fig.~\ref{fig:profiles} that shows for all
four cases typical bifurcation diagrams in dependence on $U$ and
steady height profiles for selected values of $U$: 

(a) At small $\alpha$, the volume $\Delta V$ monotonically increases:
first slowly, then faster until it diverges at about
$U_\infty\approx0.04$ [Fig.~\ref{fig:profiles}(a)]. The corresponding
simple meniscus profiles first deform only slightly due to viscous
bending before a distinguished foot-like protrusion of a height
$h_\mathrm{foot}\approx10$ develops whose length $L_\mathrm{foot}$
diverges $\propto\ln\left[(U_\infty-U)/U_\infty\right]^{-1}$.  This
corresponds to a \textit{continuous dynamic emptying transition}, a
dynamic analogue of the equilibrium transition discussed in
Ref.~\cite{PRJA2012prl}. One may also say that at $U_\infty$ the tip
of the foot unbinds from the bath. For $U>U_\infty$ the foot advances
with a constant velocity $V_F \approx (U-U_\infty)$ as shown in
Fig.~\ref{fig:frontvel}.

\begin{figure}
\includegraphics[width=0.9\hsize]{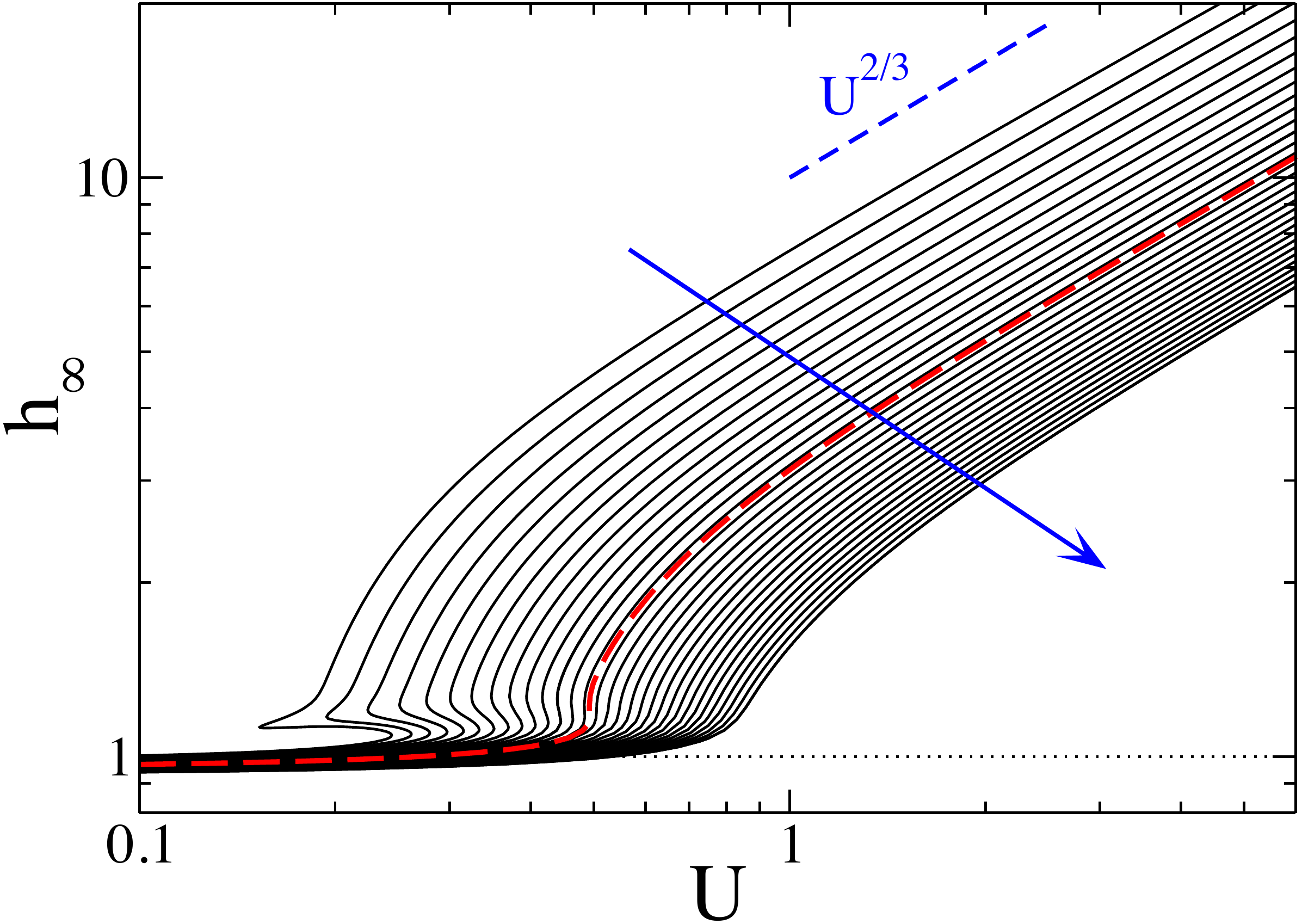}
\caption{The film height of thick (macroscopic) coatings scales
  following the Landau -- Levich law, i.e. $h_\infty\propto
  U^{2/3}$ for velocities above $U\gtrsim1$. Equidistant inclination
  angles with $\alpha\in$ [2.42, 10] and  $\Delta\alpha$ = 0.25. The
  arrow indicates increasing $\alpha$. The thick dashed line indicates
  the transition from dicontinuous to continuous dynamic wetting
  transitions occuring at $\alpha_3$.}
\label{fig:hleft_LanLev}
\end{figure}

(b) Above a first critical $\alpha=\alpha_\mathrm{1}\approx0.103$, the
transition changes its character and becomes a \textit{discontinuous
  dynamic emptying transition} that has no analogue at equilibrium. As
shown in Fig.~\ref{fig:profiles}(b), $\Delta V$ increases first
monotonically with $U$ until a saddle-node bifurcation is reached at
$U_1$ where the curve folds back. Following the curve further, one
finds that it folds again at $U_2$. This back and forth folding
infinitely continues at loci that exponentially approach $U_\infty$
from both sides and that separate linearly stable and unstable parts
of the solution branch. This exponential (or collapsed) snaking
\cite{MaBK10pd} results in foot length with
$\left[(U_\infty-U)/U_\infty\right]^{-1}\propto\exp(\mathrm{Re}[\nu]
L_\mathrm{foot})\sin(\mathrm{Im}[\nu]L_\mathrm{foot})$ where Re$[\nu]$
and Im$[\nu]$ determine the exponential approach and the period of the
snaking, respectively.  Note that for $U>U_\infty$ one can always find
a critical foot length beyond which the foot advances with a constant
velocity $V_F\approx(U-U_\infty)$. In contrast, for $U<U_\infty$ there
is always a critical length above which a foot recedes. The two cases
are illustrated in Fig.~\ref{fig:frontvel} for $\alpha=0.1$ and
$\alpha=0.5$, respectively.
In both cases, (a) and (b), one finds that the foot height
$h_\mathrm{foot}\propto U^{1/2}$. The limiting velocity
$U_\infty^\alpha$ coincides with the velocity of a large flat drop
(pancake-like drop) sliding down a resting plate of inclination
$\alpha$ \cite{TNBP2002csapea}. This allows one to calculate $U_\infty$
by continuation (see Fig.~\ref{fig:fold_vel} below). Note that the
found relation for the front velocity $V_\mathrm{F}\approx
U-U_\infty^{\alpha}$ [Fig.~\ref{fig:frontvel}(a)] is a direct
consequence of the Galilean invariance of the motion of a drop down an
incline.

(c) At a second critical $\alpha=\alpha_2\approx2.42$, the bifurcation
diagram dramatically changes. Above $\alpha_2$ the family of
steady meniscii that is connected to $U=0$ does not anymore diverge at
a limiting velocity $U_\infty$. Instead of a protruding foot of
increasing length that unbinds from the meniscus one finds a
hysteretic transition [in Fig.~\ref{fig:profiles}(c) between $U=0.1$
and $0.3$] towards a coating layer whose thickness homogeneously
increases with increasing $U$, i.e., the layer surface unbinds from
the substrate in an \textit{discontinuous dynamic wetting transition}.

\begin{figure}[htbp!]
\centering
\includegraphics[width=0.9\hsize]{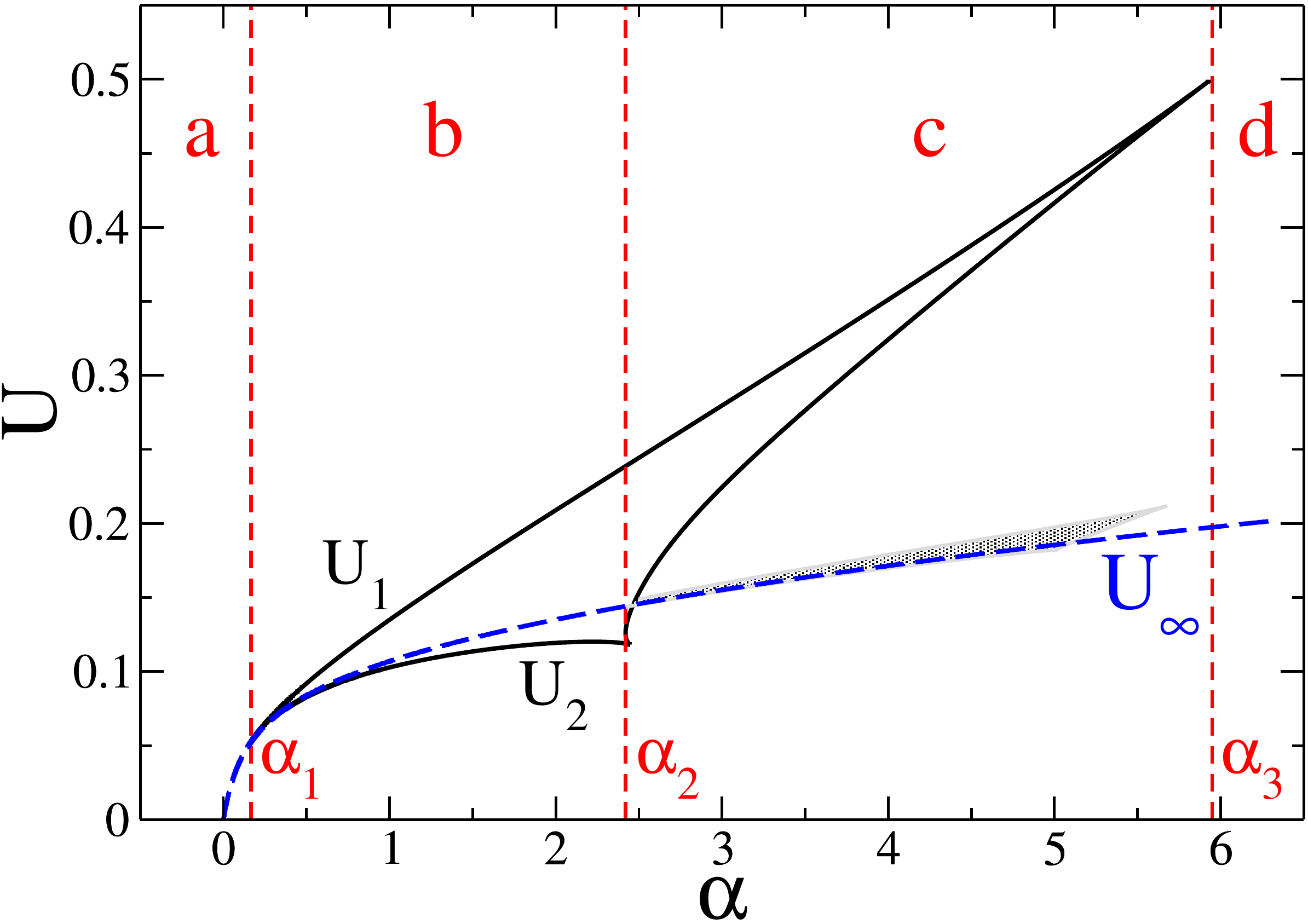}
\caption{The phase diagram in the ($U$, $\alpha$) parameter plane
  allows us to identify regions of different behaviour that are
  limited by the loci of (i) saddle-node bifurcations of steady film
  surface profiles (black solid lines) and (ii) by the dependence of
  the limiting velocity $U_\infty$ on $\alpha$ (blue dotted line). The
  existence of an additional solution family close to $U_\infty$ in
  region (c) is indicated by a grey shading. The behaviour in regions
  (a) to (d) is described in the main text. }
\label{fig:fold_vel}
\end{figure}

(d) With increasing $\alpha$ the hysteresis of the discontinuous
transition becomes smaller until at a third critical
$\alpha=\alpha_3\approx5.92$ the two saddle-node bifurcations
annihilate in a hysteresis bifurcation as further illustrated in
Fig.~\ref{fig:hleft_LanLev}. For all $\alpha>\alpha_3$ one finds a
\textit{continuous dynamic wetting transition}.  As in both cases -
(c) and (d) - at large $U$ the coating layer thickness follows the
power law $h_\infty\propto U^{2/3}$, we identify these unbinding
states as Landau-Levich films \cite{LaLe1942apu}. The critical
velocity where the transition between the microscopic and macroscopic
layer occurs, scales as $\alpha^{3/2}$.

Note that the dynamic emptying transitions of cases (a) and (b) and
the crossover between them is also observed employing a slip model
\cite{SADF2007jfm,ZiSE2009epjt}. However, as normally slip models do
not take account of the mesoscale wetting behaviour they are unable to
describe the discontinuous and continuous dynamic wetting transitions
of cases (c) and (d), respectively, as these represent transitions
between the topologically different meniscus and film solutions.

To summarize our findings we present in Fig.~\ref{fig:fold_vel} a
phase diagram in the plane spanned by the plate velocity and
inclination angle as obtained by tracking the main occuring
saddle-node bifurcations (black solid lines) and the limiting velocity
$U_\infty$ (blue dotted line).
In region (a), i.e., for $0<\alpha< \alpha_1$, ultimately a simple or
extended (foot-like) steady meniscus is found for $U<U_\infty$ while
for $U>U_\infty$ the foot advances at constant speed $V_\mathrm{F}\approx
(U-U_\infty)$ [cf.~Fig.~\ref{fig:frontvel}(a)].
In region (b), i.e., for $\alpha_1<\alpha< \alpha_2$, multiple stable
foot solutions exist for $U$ between the two solid lines. However, for
each $U$ with $U_2>U>U_\infty$ there is always a maximal stable foot
length $L_\mathrm{max}^-$ towards which a longer foot will
retract. For each $U$ with $U_\infty>U>U_1$ there is always a maximal
unstable steady foot length $L_\mathrm{max}^+$ beyond which the foot
will prolongate continuously.  $L_\mathrm{max}^-$ [$L_\mathrm{max}^+$]
logarithmically diverges as $U$ approaches $U_\infty$ from below
[above].

\begin{figure}
\includegraphics[width=1\hsize]{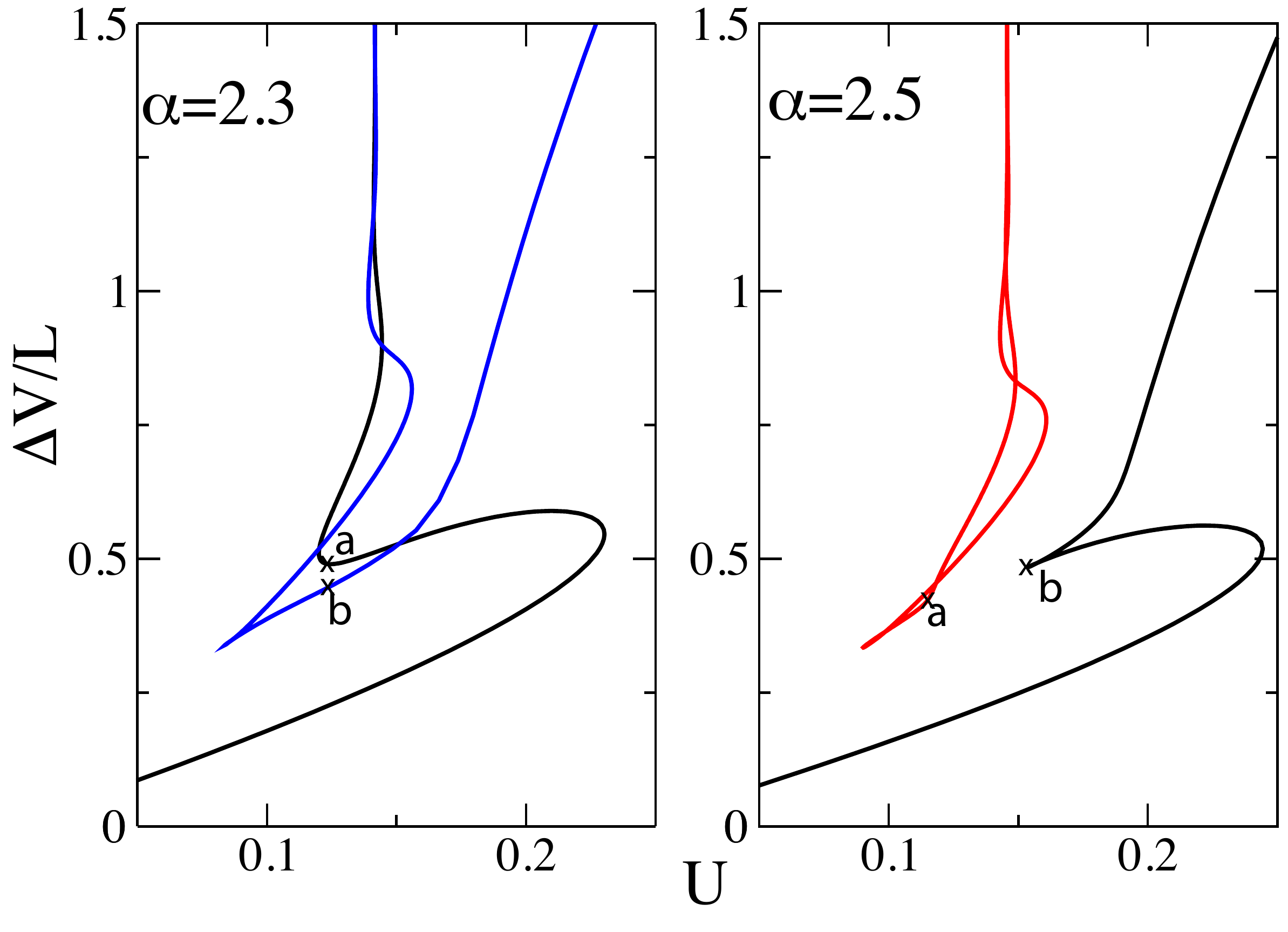}
\caption{Detail of the transition and full bifurcation diagram gathering the two families of solutions. Observe that the transition occurs via reverse--necking bifurcation. Branches are reconnecting at the marked points in the diagrams. The reconnection occurs at $\alpha=\alpha_2$. The domain size is $L = 1000$.}
\label{fig:transition_volb}
\end{figure}
In region (c), i.e., for $\alpha_2<\alpha< \alpha_3$,  the lines of
saddle-node bifurcations limit a region where initial conditions
decide whether an ultrathin layer or a macroscopic Landau-Levich 
coating is obtained. Below [above] the hysteresis range one only finds
the ultrathin [Landau-Levich] coating. 
In region (d), i.e., for $\alpha_3<\alpha$, the change between the two
coating types is continuous.

Although an extended analysis of the characteristics of the observed
qualitative changes with increasing inclination angle is beyond our
present scope, we highlight some further important facts. The crossover
between regions (a) and (b) at $\alpha=\alpha_1$ can be understood in terms
of a change of the character of the spatial eigenvalues (EV) of a flat
film of a height that corresponds to the foot height
\cite{ZiSE2009epjt,TsGT2013arxiv}: In region (a) all EV are real while
in region (b) only one is real and the other two are a pair of complex
conjugate EV. The crossover between regions (c) and (d) at $\alpha=\alpha_3$
results from a hysteresis bifurcation where two saddle-node
bifurcations annihilate. However, the crossover between regions (b)
and (c) at $\alpha=\alpha_2$ that results in the strongest qualitative
change, namely, from a dynamic emptying to a dynamic wetting
transition cannot be understood by analysing a single family of steady
profiles. As illustrated in Fig.~\ref{fig:transition_volb} the
crossover results from a reconnection (reverse necking bifurcation) at
$\alpha=\alpha_2$ that involves two solution families. Both
continue to exist on both sides of $\alpha_2$. This results in
intricate behaviour in certain small bands of the ($U$, $\alpha$)
plane and, in particular, around $\alpha_2$ that will be studied in
more depth elsewhere. For instance, in the fine grey band around
$U_\infty$ in region (c) [Fig.~\ref{fig:fold_vel}], there exist
various stable extended meniscus profiles. They correspond to the left
branch in Fig.~\ref{fig:transition_volb}(b).  Experimentally, they
might only be obtained through a careful control of the set-up at
specific initial conditions.

To conclude, we have shown that a long-wave mesoscopic hydrodynamic
description of the coating problem for a plate that is drawn from a
bath allows one to identify and analyze several qualitative
transitions if wettability is modelled via a Derjaguin pressure, i.e.,
with a precursor film model.  As a result we have distinguished four
regions where different dynamic unbinding transitions occur, namely
continuous and discontinuous dynamic emptying transitions and
discontinuous and continuous dynamic wetting transitions. These
dynamic transitions are out-of-equilibrium equivalents of well known
equilibrium emptying and wetting transitions. Beside features known
from the equilibrium versions, our analysis has uncovered several
important features that have no equivalents at equilibrium. A future
study of the influence of fluctuations might allow one to answer the
question which surface profile is selected in the multistable regions.

We acknowledge support by the EU via the FP7 Marie Curie scheme (ITN
MULTIFLOW, PITN-GA-2008-214919).

\renewcommand*{\refname}{}
\bibliographystyle{unsrt}
\bibliography{GTLT13}

\begin{thebibliography}{10}
\bibitem{noteScaling}
The employed scales are $\ell=\sqrt{3/5} h_\mathrm{eq}/\theta_\mathrm{eq}$,
  $h_\mathrm{eq}$ and $\tau=(3\eta
  h_\mathrm{eq})/(25\gamma\theta^{4}_\mathrm{eq})$ for the $x$-coordinate, film
  height and time, respectively. For the dimensional disjoining pressure
  $\Pi=-A/h^3 + B/h^6$, with Hamaker constant $A>0$ and short-range interaction
  strength $B>0$, the equilibrium precursor height and contact angle are
  $h_\mathrm{eq}=(B/A)^{1/3}$ (i.e., the scaled precursor height is $h_p=1$)
  and $\theta_\mathrm{eq}=\sqrt{3A/5\gamma h^2_\mathrm{eq}}$ (i.e., the scaled
  equilibrium contact angle is $\sqrt{3/5}$), respectively. The velocity scale,
  gravity number and the scaled inclination angle are given by $3\tau/\ell$,
  $G=\rho g h_\mathrm{eq}^4/A$, and $\alpha=(\ell/h_\mathrm{eq})\tilde
  \alpha=O(1)$, respectively, where $\rho$ is the density of the liquid, $g$ is
  gravitational accelleration and $\tilde \alpha$ is the physical plate
  inclination angle. We use $G=0.001$.
\bibitem{notehamiltonian}
Here, $\xi dx\approx (1/2)[1+(\partial_x h)^2]ds$ is the surface area element
  in long-wave approximation and $f(h)$ is an appropriately defined energy
  containing terms related to wettability and gravity. Note that in the wetting
  literature $F[h]$ is called "Hamiltonian" as it may be derived from a
  microscopic Hamiltonian. However, technically it is a Lagrangian as the
  steady version of Eq.~(\ref{eq:thifi_dim_non}) for horizontal resting
  substrate (without gravity) corresponds to the Euler-Lagrange equation of
  $F[h]$ under the condition of mass conservation.
\bibitem{notefoot}
We define the measure for the foot length $L_\mathrm{foot}\propto\Delta
  V/(h_\mathrm{f}-h_\infty)$, where $h_\mathrm{f}$ is the foot height and
  $h_\infty$ the coating film height.
\end{thebibliography}
\end{document}